\documentclass[pre,aps,showpacs,twocolumn]{revtex4-1}
\usepackage{bm}
\usepackage{amsmath,amssymb,amsfonts,latexsym,fancyhdr,graphicx,epstopdf,times,txfonts}
\usepackage{caption}
\usepackage{subcaption}
\usepackage{lipsum}
\begin{document}
\title{Fluctuation relations for driven coupled classical two-level systems}
\author{Massimo Borrelli$^1$, Jonne V. Koski$^{1}$, Sabrina Maniscalco$^2$, and Jukka P. Pekola$^{1}$}
\affiliation{$^1$ Low Temperature Laboratory (OVLL), Aalto University, POB 13500, FI-00076 AALTO, Finland\\
$^2$ Turku Centre for Quantum Physics, Department of Physics and Astronomy, University of
Turku, FI-20014 Turun yliopisto, Finland\\}

\begin{abstract}
We theoretically investigate fluctuation relations in a classical incomplete measurement process where just partial information is available. The scenario we consider consists of two coupled single-electron boxes where one or both devices can undergo a non-equilibrium transformation according to a chosen protocol. The entropy production of only one of the two boxes is recorded and fluctuation relations for this quantity are put to a test, showing strong modifications whose nature depends upon the specific case study.  
\end{abstract}

\maketitle

\section{Introduction}
Work fluctuation relations link the thermodynamic behavior of a system undergoing a non-equilibrium transformation to its equilibrium properties \cite{evans,gavallotti,seifertrev}. The most prominent examples of such laws are the Jarzynski equality \cite{jarz1,jarz2} and Crooks relation \cite{crooks}. These relations  have been experimentally confirmed in several setups such as colloidal particles \cite{wang,blickle}, biological molecules \cite{dna}, defects in diamonds \cite{schuler} and electronic nanostructures \cite{olli,jonne}. Moreover, these experimental achievements have stimulated a great deal of further theoretical studies \cite{seifert1,sagawafeed,esposito,aki,masked}. The typical scenario is the following. The system under scrutiny is embedded in a thermal environment and driven between two different Hamiltonians over a time that is short enough not to allow instantaneous equilibration. The exponentiated work is then averaged over many repetitions of the same protocol and equilibrium free energy differences can be extracted. 
In all of these cases a complete identification of the relevant dynamical quantities is required. Thus, it is only natural to wonder what kind of non-equilibrium statistics we would observe if such an assumption no longer held true. This problem has been experimentally addressed \cite{mehl} in order to understand how the lack of knowledge of slow degrees of freedom in a non-equilibrium system might modify standard fluctuation relations. Two colloidal particles were forced to interact by switching on and off a static magnetic field and the dynamics of only one particle was tracked. For most of the experimental parameters as well as for small and large values of the entropy production a Crooks-type relation \cite{evans} was observed
\begin{equation}
\log\left[\frac{P(\Delta S)}{P(-\Delta S)}\right]=\alpha\Delta S,
\label{eq1}
\end{equation}
where, however, the slope $\alpha$ deviated from unity.\\
Here, we take a different but complementary approach. First of all, the physical setup we consider is a pair of coupled single-electron boxes (SEB) \cite{averin,buttiker,lafarge}. These are electronic nano-circuits where single-electron currents can be generated and controlled. Although the basic mechanism inducing such currents is quantum tunneling of single electrons across a junction, we assume our stochastic system to operate in the classical domain. In other words, no quantum coherences between different electronic states are present at any time. This regime has been experimentally implemented in a series of recent works  \cite{olli,jonne,demon,engine}. Here non-equilibrium dynamics has been observed by implementing suitable time-dependent protocols changing some relevant energy parameters, such as voltage. We investigate two possible scenarios. While the two SEBs interact through a time-independent force we drive the single-box energy of either one or both the boxes according to a fixed time-dependent protocol. 
We discard completely the dynamics of the other box, say 2, and calculate the entropy production in box 1. The main question we address is the following: how does such an interaction between the two SEBs affect the driven dynamics and consequently the non-equilibrium statistics of the single box? Deviations from the standard behavior are found and, although their extent as well as their nature depend upon the specific details of the driving protocol, a general trend arises. In all of the situations considered such deviations are non-linear both in the coupling strength and in the entropy production. Thus, they are not explainable with an effective-thermal-environment description for the reduced dynamics and statistics of box 1. Again, this is a complementary viewpoint to the approach adopted in \cite{mehl} as we imagine to drive the free Hamiltonian/s of one or both the SEBs while the interaction is always present although static. One can look at this model as the prototype of an incomplete experimental setup where information regarding some degrees of freedom is missing. In this respect, this study aims at a better understanding of which degrees of freedom and time-scales are truly relevant when it comes to non-equilibrium physics, with particular emphasis on fluctuation relations. 
This article is organized as follows: in Sec.~\ref{model} we will introduce and describe our model; in Sec.~\ref{therm} we will review some basic facts regarding fluctuation relations; in Sec.~\ref{single} and Sec.~\ref{double} we will present our findings, and the final section will be devoted to conclusions.
\section{The model}\label{model}
\begin{figure}
\centering
\begin{subfigure}{0.5\textwidth}
\includegraphics[width=\textwidth]{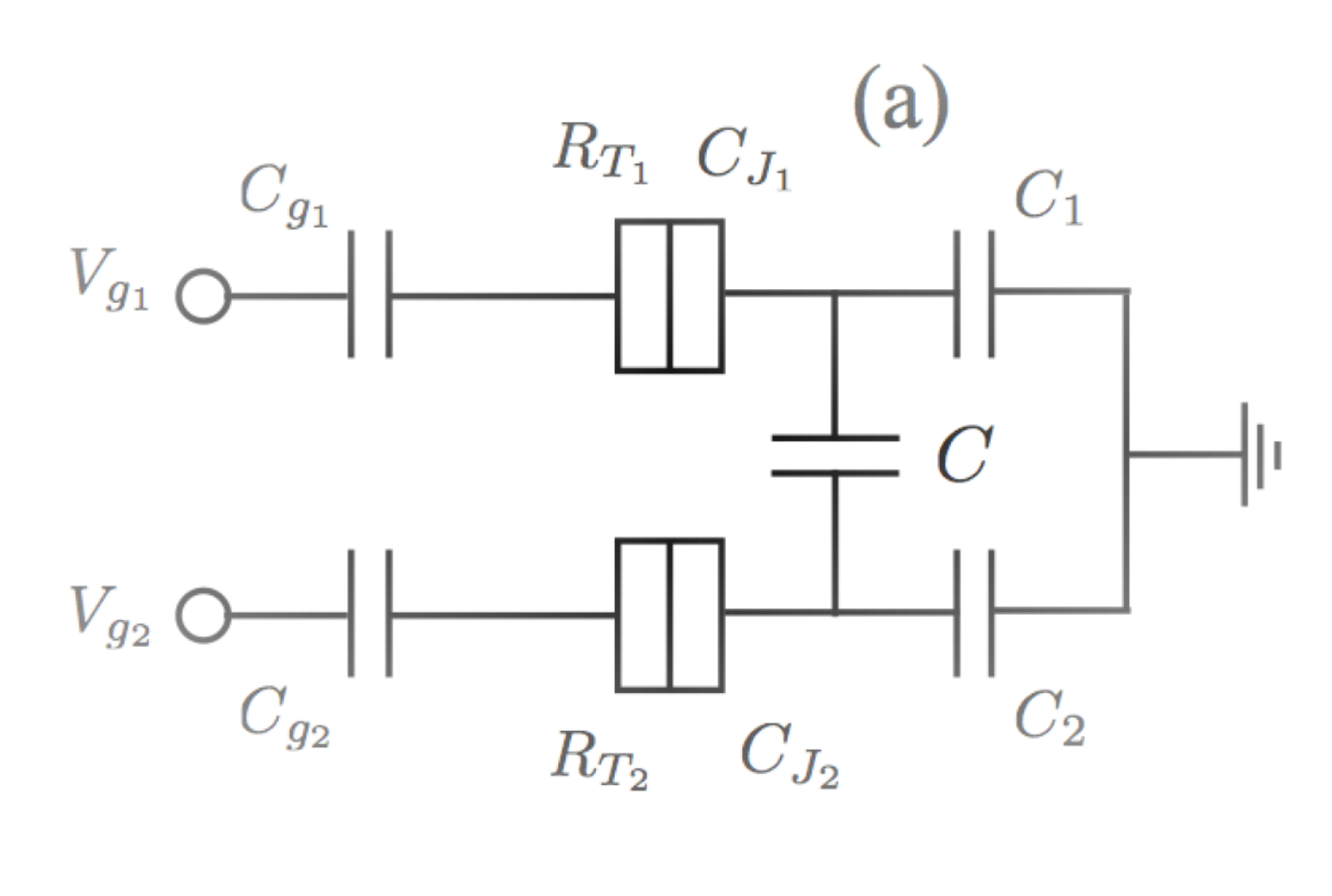}
\end{subfigure}
\begin{subfigure}{0.239\textwidth}
\includegraphics[width=\textwidth]{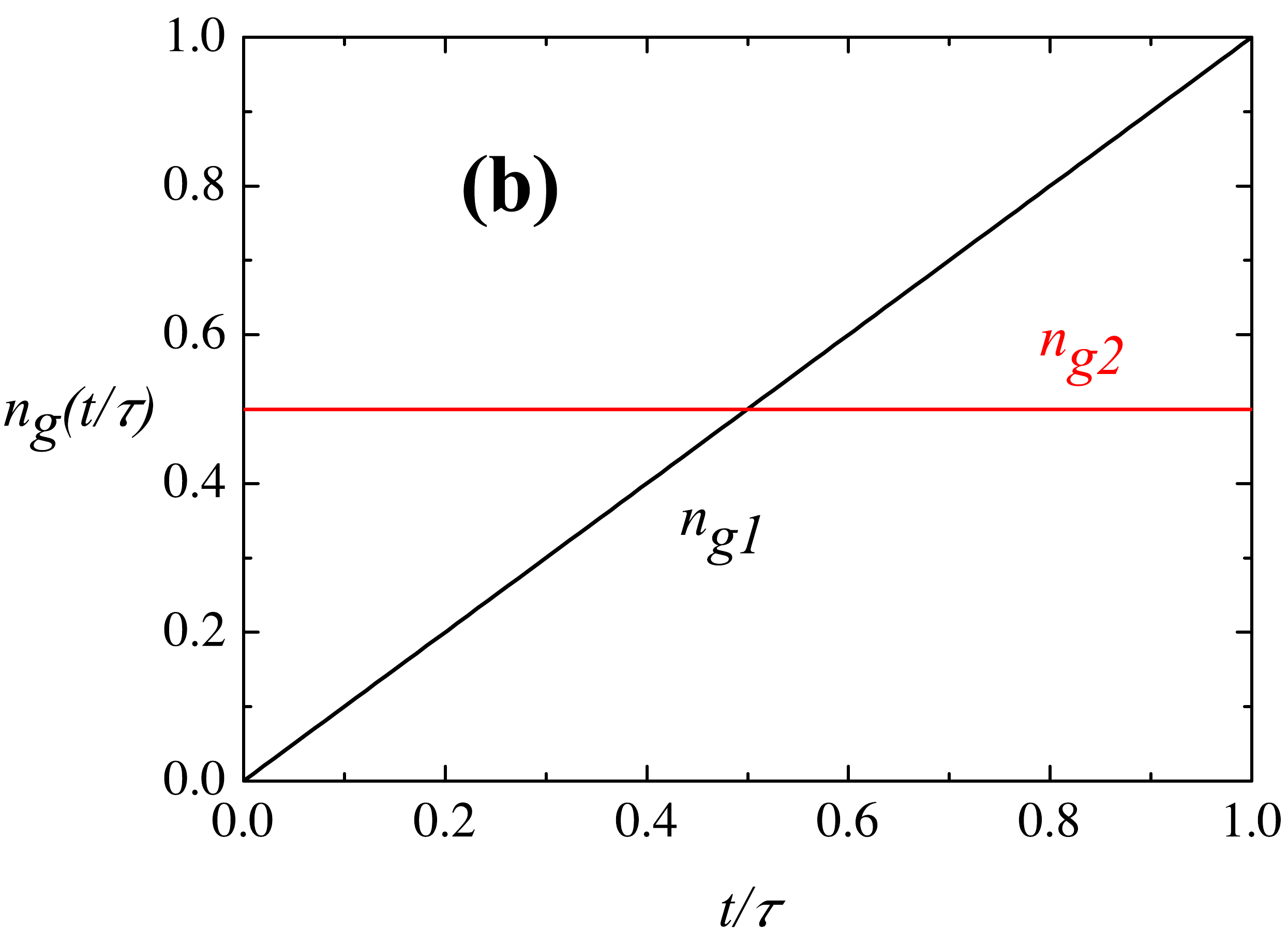}
\end{subfigure}
\begin{subfigure}{0.239\textwidth}
\includegraphics[width=\textwidth]{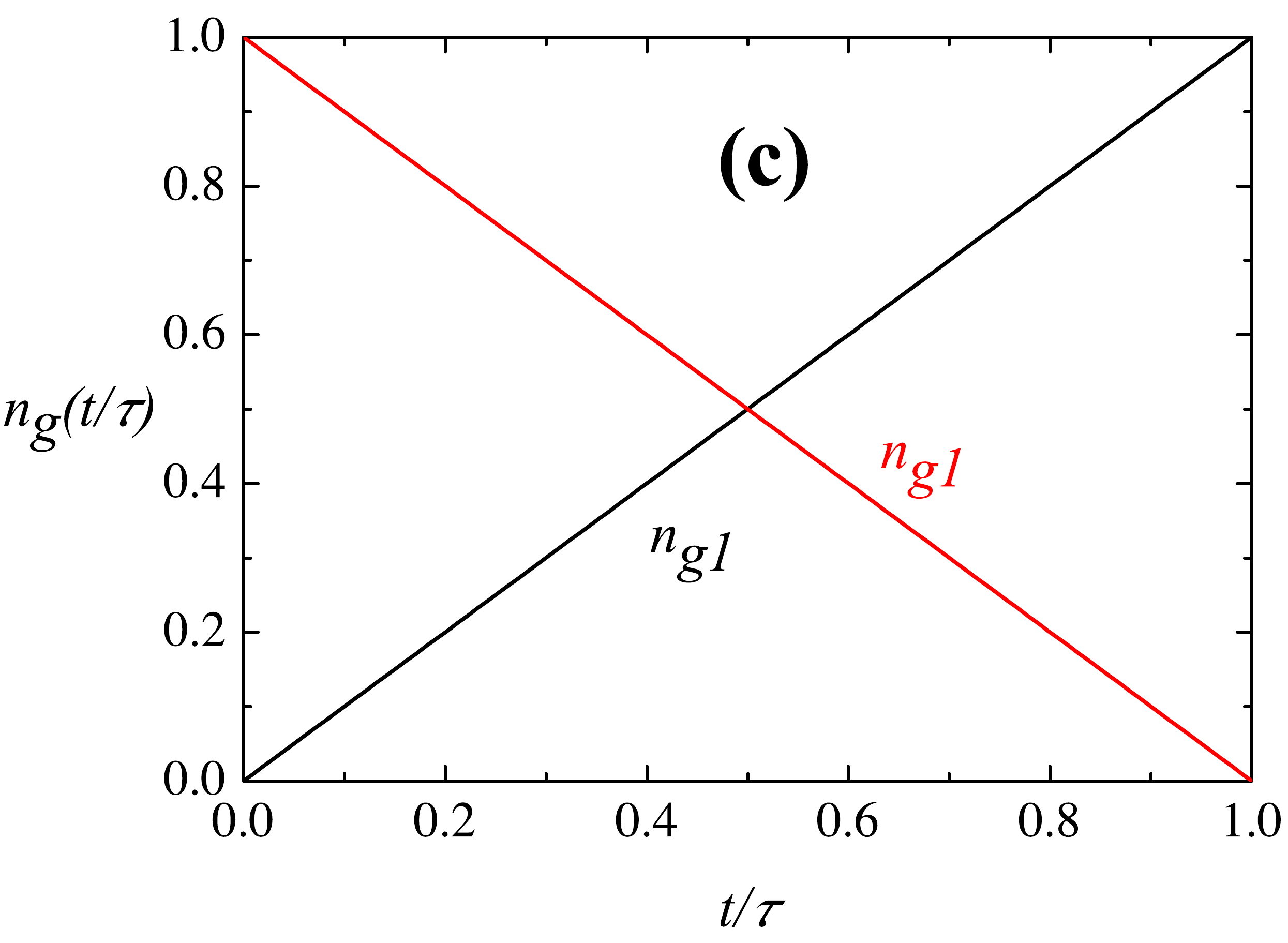}
\end{subfigure}
\caption{(a) Circuital implementation of coupled single-electron boxes satisfying Eq.~\eqref{hamiltonian}; the divided box represents the tunnel junction through which electrons tunnel by rates \eqref{rategeneral}. (b) First protocol implemented where the first box is linearly driven from $n_{g_{1}}=0$ to $n_{g_{1}}=1$ (black) while the second box is kept at $n_{g_{2}}=1/2$ (red). (c) Second protocol implemented where the two boxes are driven between opposite ground states, $n_{g_{1}}:0\to1$ (black) and $n_{g_{2}}:1\to0$ (red).}
\label{fig1}
\end{figure}

We consider two single-electron boxes capacitively interacting with each other, see Fig.\ref{fig1}a. The non-equilibrium thermodynamics of these nano-devices has recently been the subject of intense theoretical and experimental investigations \cite{jukka1,jukka2,jukka3}. Each box consists of two conducting electrodes coupled through a junction with capacitance $C_{J_{i}}, i=1,2$  and biased by a gate voltage $V_{g_{i}}$ applied through a capacitor $C_{g_{i}}$. The total system is at thermal equilibrium with the surroundings at temperature $\beta^{-1}=k_{B}T$. The Hamiltonian governing the dynamics reads
\begin{equation}
H=E_{C_{1}}(n_{1}-n_{g_{1}})^{2}+E_{C_{2}}(n_{2}-n_{g_{2}})^{2}+J(n_{1}-n_{g_{1}})(n_{2}-n_{g_{2}}),
\label{hamiltonian}
\end{equation}
where $E_{C_{i}}=e^{2}/(2C_{\Sigma_{i}})$ is the $i^{\textrm{th}}$ single-box charging energy, with $C_{\Sigma_{i}}=C_{J_{i}}+C_{g_{i}}$ being the total capacitance, $n_{i}$ is the number of excess electrons, $n_{g_{i}}=-C_{g_{i}}V_{g_{i}}/e$ is the charge of the gate voltage, and $J=e^{2}C/[C_{\Sigma_{1}}C_{\Sigma_{2}}+C(C_{\Sigma_{1}}+C_{\Sigma_{2}})]$ is the intra-box coupling constant. Changing the gate voltage of a box causes excess electrons to tunnel across the junction. The variables $n_{i}$ are therefore stochastic, integer valued and unbounded. For most of the applications the surrounding temperature can be made small enough to guarantee that $\beta E_{C}\gg1$. Furthermore, if the resistance of each junction $R_{T}$ is tuned such that $R_{T}^{-1}\ll e^{2}/\hbar$, only the two lowest electron states $n=0,1$ can be effectively populated. With this simplification only four two-box states need to be accounted for, {\it i.e.} $(n_{1},n_{2})=(0,0), (1,0), (0,1), (1,1)$. Depending on the value of the SEB resistance, both single and many-electron co-tunneling processes can take place. From now on we will assume only single-electron processes to be  relevant. This means that only the following transitions are to be considered
\begin{equation}\begin{aligned}
&(0,0)\leftrightarrow(1,0)\\
&(0,0)\leftrightarrow(0,1)\\
&(1,0)\leftrightarrow(1,1)\\
&(0,1)\leftrightarrow(1,1).
\label{trans}
\end{aligned}
\end{equation}
and the energy differences for the corresponding transitions are
\begin{equation}\begin{aligned}
&\Delta E_{(0,0)\leftrightarrow(1,0)}=\pm(E_{C_{1}}-2E_{C_{1}}n_{g_{1}}-Jn_{g_{2}})\\
&\Delta E_{(0,0)\leftrightarrow(0,1)}=\pm(E_{C_{2}}-2E_{C_{2}}n_{g_{2}}-Jn_{g_{1}})\\
&\Delta E_{(1,0)\leftrightarrow(1,1)}=\pm[E_{C_{2}}+ J(1-n_{g_{1}}) - 2E_{C_{2}}n_{g_{2}}]\\
&\Delta E_{(0,1)\leftrightarrow(1,1)}=\pm[E_{C_{1}}+ J(1-n_{g_{2}}) - 2E_{C_{1}}n_{g_{1}}].
\label{entrans}
\end{aligned}
\end{equation}
The time-evolution of the occupation probabilities $p_{n_{1}n_{2}}(t)$ is governed by a system of rate equations with a time-dependent transition-rate matrix $A(t)$
\begin{equation}
\dot{\vec{p}}(t)=A(t)\vec{p}(t),
\label{RE}
\end{equation}
where $\vec{p}(t)$ is the occupation probability vector and the elements of $A(t)$ are the transition rates. For a transition corresponding to an energy difference $\Delta E$ the corresponding rate reads as 
\begin{equation}
\Gamma(\Delta E)=\frac{1}{R_{T}e^{2}}\frac{\Delta E}{e^{\beta\Delta E}-1}.
\label{rategeneral}
\end{equation}
These rates will be also used to generate the stochastic trajectories.

\section{Non-equilibrium fluctuation relations}\label{therm}
We focus on two well-known fluctuation relations. The first one is the Jarzynski equality \cite{jarz1,jarz2}. This links the work $W$ performed in a general thermodynamic transformation to equilibrium free energy difference $\Delta F$ for a system in a single bath. Shortly, the Hamiltonian $H_{i}$ of a system initially at thermal equilibrium is changed in time to a final $H_{f}$, for instance by driving one of its parameters according to some protocol. No assumption regarding the duration of such a protocol is made. The work performed $W$ is recorded. If $\Delta F=-\log(Z_{f}/Z_{i})$ is the equilibrium free energy difference between the initial and final Boltzmann distributions, the following relation holds
\begin{equation}
\langle e^{-\beta W}\rangle=e^{-\beta\Delta F},
\label{Jarz}
\end{equation}  
where $\langle\cdots\rangle$ denotes a statistical average over many repetitions of the same protocol.
In the model utilized here and for all the protocols we consider, the equilibrium free energy difference $\Delta F$ vanishes exactly. Hence, by using the first law of thermodynamics, the Jarzynski relation can be recast in the following form
\begin{equation}
\langle e^{-\Delta S}\rangle=1,
\label{JarzQ}
\end{equation}
where $\Delta S$ is the total entropy production. The heat dissipated to the bath in a single run of the protocol can be shown to be proportional to the statistical entropy production $\Delta S$ \cite{jonne}
\begin{equation}
Q=-\beta\Delta S=-\beta\sum_{j}\log\left[\frac{\Gamma(\Delta E(t_{j}))}{\Gamma(-\Delta E(t_{j}))}\right],
\label{Q}
\end{equation}
where a transition associated to the energy gap $\Delta E$ occurs at time $t_{j}$. We remind that the above expression refers to the total system and to all the possible single-electron transitions that occur in both single-electron boxes. 
The second relation we focus on is the fluctuation theorem \cite{evans}
\begin{equation} 
\log\left[\frac{P(\Delta S)}{P(-\Delta S)}\right]=\Delta S,
\label{totalTC}
\end{equation}
where $P(\Delta S) (P(-\Delta S))$ is the entropy production probability distribution of the forward (backward) transformation. 
The question we address is whether ignoring some degrees of freedom may lead to deviations from Eqs.\eqref{JarzQ}-\eqref{totalTC}. 
We are going to be investigate the statistics of stochastic entropy production $\Delta S_{1}$ in the first box
\begin{equation}
\Delta S_{1}=\sum_{j\in\mathcal{B}_{1}}\log\left[\frac{\Gamma(\Delta E(t_{j}))}{\Gamma(-\Delta E(t_{j}))}\right],
\label{Entropy1}
\end{equation}
where $\mathcal{B}_{1}$ represents the subset of transitions occurring in the first SEB only. The idea is then to investigate the following quantities
\begin{equation} 
\log\left[\frac{P(\Delta S_{1})}{P(-\Delta S_{1})}\right],
\label{TC1}
\end{equation}
\begin{equation}
\langle e^{-\Delta S_{1}}\rangle.
\label{Jarz1}
\end{equation}

\section{Single drive}\label{single}
In this section we focus on the following scenario. We are given a single-electron box, capacitively coupled to a second box whose existence we are not aware of. We change the gate voltage $n_{g_{1}}$ of the first SEB in time as follows
\begin{equation}
n_{g_{1}}(t)=\frac{t}{\tau},
\label{drive1}
\end{equation}
from $t=0$ to $t=\tau$, see Fig.\ref{fig1}b. This will force the box to switch from the state $n_{1}=0$ to the state $n_{1}=1$. The gate voltage of the second box is instead constant $n_{g_{2}}=1/2$ at all times. This choice guarantees that $\Delta F=0$. Even though the second box is not externally driven, the interaction in Eq.~\eqref{hamiltonian} may cause tunneling events in the second box. Thus, contributions from box 2 to the thermodynamics of the total system will arise. Our goal is to investigate whether ignoring such contributions results in a modified version of standard fluctuation relations, such as Eqs.~\eqref{TC1}-\eqref{Jarz1}. Summarizing, the experimental scenario considered here is the following: 1) the initial two-box state is described by a Boltzmann distribution $\exp(-\beta H)$ where $H$ is given in  Eq.~\eqref{hamiltonian}; 2) we change the gate voltage of the first box linearly in time according to Eq.~\eqref{drive1} while keeping the gate voltage of the second box at $n_{g_{2}}=1/2$; 3) we generate a trajectory for the joint stochastic variables $(n_{1}(t), n_{2}(t))$ using Monte-Carlo jump method; 4) we record the entropy $\Delta S_{1}$ generated in the first box only; 5) based on the outcomes of each single repetition we perform a statistical analysis of Eqs.~\eqref{JarzQ}-\eqref{totalTC}, but for $\Delta S_{1}$ instead of $\Delta S$. We anticipate that deviations are observed. However, when the statistics of $\Delta S$ is considered, both Eq.~\eqref{JarzQ} and Eq.~\eqref{totalTC} are recovered.

\subsection{Identical SEBs}
\begin{figure}
\centering
\includegraphics[scale=0.3]{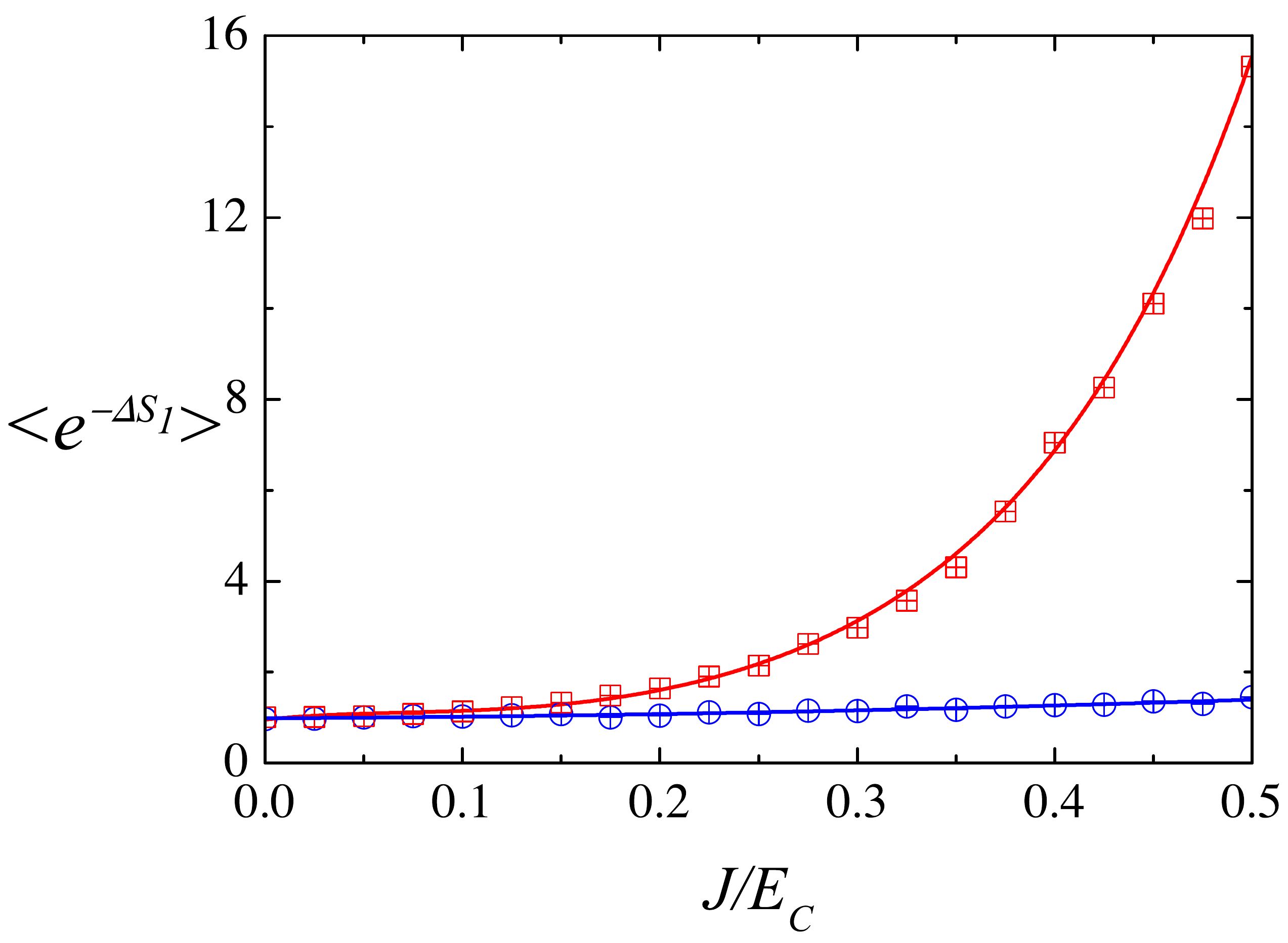}
\caption{Average exponentiated reduced entropy production $\langle e^{-\Delta S_{1}}\rangle$ as a function of the relative interaction strength $J/E_{C}$ for two different values of the protocol duration: $\tau\Gamma_{0}=94$ (red) and $\tau\Gamma_{0}=9.4$ (blue) The continuous lines are quadratic fits.}
\label{plot1}
\end{figure}

\begin{figure}
\centering
\begin{subfigure}{0.4\textwidth}
\includegraphics[width=\textwidth]{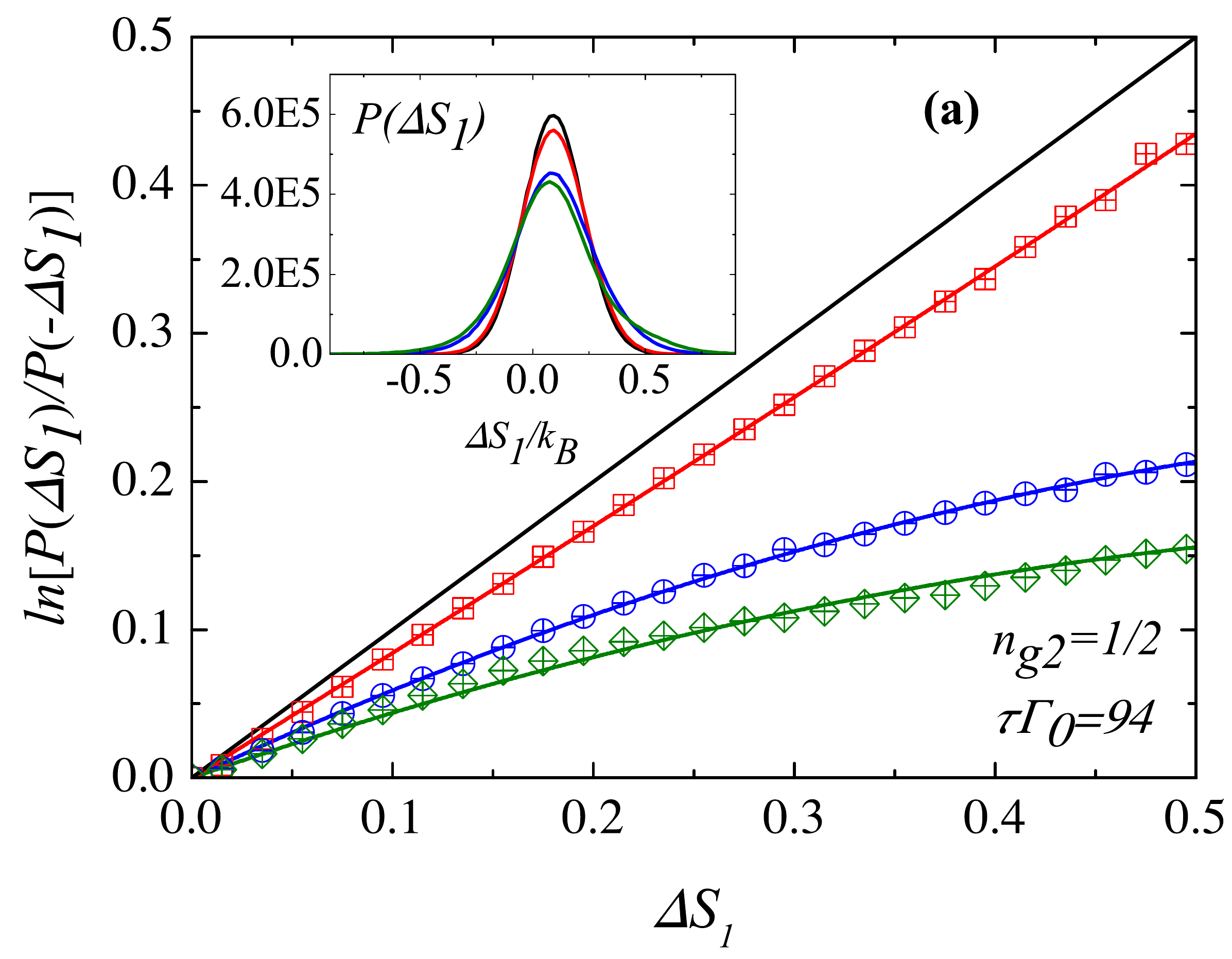}
\end{subfigure}
\begin{subfigure}{0.4\textwidth}
\centering
\includegraphics[width=\textwidth]{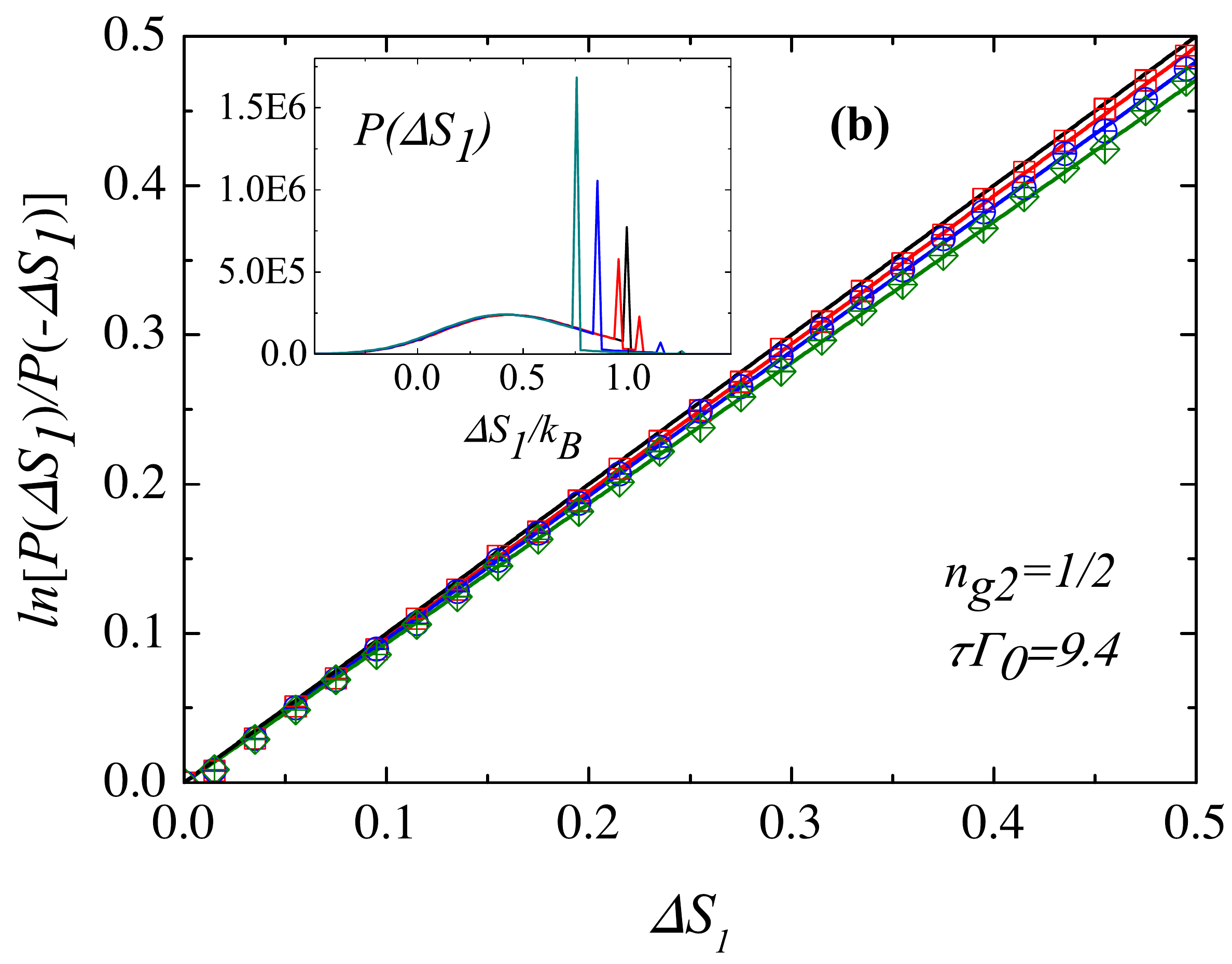}
\end{subfigure}
\caption{Crooks-type fluctuation relation for the single-box entropy production $\Delta S_1/k_{B}$ in the case of a slow (a) and fast (b) drive, with $\tau\Gamma_0 =94,\;9.4$ respectively. Different colors correspond to $J/E_{C}=0.1$ (red), $J/E_{C}=0.3$ (blue), $J/E_{C}=0.5$ (dark green). The black continuous line refers to $J/E_{C}=0$ and it is displayed for completeness. The symbols are the values obtained from numerical simulations while the continuous lines are extracted by a quadratic fitting of these values. Inset: probability distribution of the single-box entropy production $P(\Delta S_{1})$.}
\label{plot2}
\end{figure}

Here we assume the two single-electron boxes to be identical ($E_{C_{1}}=E_{C_{2}}=E_{C}, R_{T_{1}}=R_{T_{2}}=R_{T}$). 
We define $E_{C}/R_{T}e^{2}\equiv\Gamma_{0}$ and choose two different durations of the driving protocol $\tau\Gamma_{0}=9.4,94$. The temperature will be chosen at $\beta E_{C}=10$, a value used in all the numerical examples here.
In a real experimental setup, such as the one utilized in \cite{jonne}, these values would roughly correspond to $E_{C}=1.9 k_{B}\; \textrm{K}, T=180 \textrm{mK}$.
The number of stochastic trajectories generated for each simulated experiment is 10 millions.\\
Figure~\ref{plot1} shows the behavior of $\langle e^{-\Delta S_{1}}\rangle$ as a function of the rescaled interaction strength  $J/E_{C}$ for two different values of $\tau$. The symbols are raw data from numerical simulations while the continuous lines are quadratic fits with a 0.95 statistical confidence level. We notice a dramatic deviation from unity in the case of long ($\tau\Gamma_0 =94$) protocol. Generally speaking, the exponentiated entropy generated in the first box is an increasing function of the coupling constant. In the case of a short protocol ($\tau\Gamma_0 =9.4$) such behavior is instead slightly harder to observe, especially for very small values of $J/E_{C}$. This result can be understood as follows. When the length of the driving protocol is long as compared to the typical time-scale of the dissipative dynamics, set by $\Gamma_{0}$, it is more likely for single-electron transitions in the second box to take place. These can be seen as a response of the second box to tunneling events occurring in the first box. A slower rate of change $1/\tau$ will allow the second box to follow the dynamics of the first one and react to it through the interaction term. A faster rate of change will instead cause the second box to essentially freeze since tunneling events in the first box occur on a much shorter time-scale. In other words, only the driven box is responsible for almost all of the entropy generated. Obviously, the more transitions in box 2 take place, the more they will contribute to the entropy production of the total system. On the contrary, if the largest contribution to the thermodynamics comes from box 1 only, neglecting box 2 will not influence much and the standard fluctuation relation is approached.\\
In Fig.~\ref{plot2}a we plot the left-hand-side of Eq.~\eqref{TC1} as a function of $\Delta S_{1}$, with $k_{B}$ being the Boltzmann constant, for increasing values of the box-box coupling constant $J$ and for a slow drive with $\tau\Gamma_0 =94$. The symbols represent the raw numerical values obtained from Monte-Carlo simulations while the continuous lines are quadratic fittings with a statistical confidence level of 0.95. While for $J=0$ we recover the standard fluctuation theorem \eqref{totalTC}, stronger deviations for progressively larger values of $J$ can be seen.  Interestingly, for stronger couplings and relatively large values of $\Delta S_{1}$ the non-linear behavior arises. The total thermal environment surrounding the two SEBs and the second box do not act as an effective thermal environment for the first box. In other words, no effective temperature for the first box exists.\\ 
Figure~\ref{plot2}b shows the same quantity in the case of a fast driving protocol with $\tau\Gamma_0 =9.4$. Again, we show both the raw numerical data and their quadratic fittings. 
 Here, deviations from the standard Eq.~\eqref{totalTC} are harder to observe. Even for large values of box 1 entropy production and the strongest coupling considered $J/E_{C}=0.5$, all the curves appear to depart very little from a line with slope 1 and non-linear contributions are not prominent. These results are in agreement with Fig.~\ref{plot1} and can be understood with the same argument. Both in Fig.~\ref{plot2}a and \ref{plot2}b we show the probability distributions $P(\Delta S_{1})$ corresponding to the different values of $J/E_C$ (inset). While In the case of a slow protocol we observe a smooth, Gaussian-type distribution for every value of the coupling, the situation is different in the case of a short drive. For all the values of the interaction strength, $P(\Delta S_{1})$ displays one or more peaks. This can be understood as follows. When $n_{g_{2}}=1/2$ the initial energy of the states $(0,0), (0,1)$ is equal, leading to a $50\%$ population probability in the initial thermal distribution $\exp(-\beta H)$ for any value of $J$. Since in this case the protocol is very fast, often no transitions will occur. Formally, $(0,0)\rightarrow(0,0)$ and $(0,1)\rightarrow(0,1)$  with $H(0,0;t=\tau)-H(0,0;t=0)=H(0,0;t=\tau)-H(0,1;t=0)=E_C$. When instead $J\ne0$ these energy differences split with a  gap $J$, leading to two distinct peaks. However, as $J$ increases the peak at $\Delta S_{1}=-1-J/2$, corresponding to $(0,0)\rightarrow(0,0)$, becomes progressively shorter since this no-transition realization is energetically less favorable than $(0,1)\rightarrow(0,1)$.  
\subsection{Unequal SEBs}

\begin{figure}
\centering
\begin{subfigure}{0.4\textwidth}
\includegraphics[width=\textwidth]{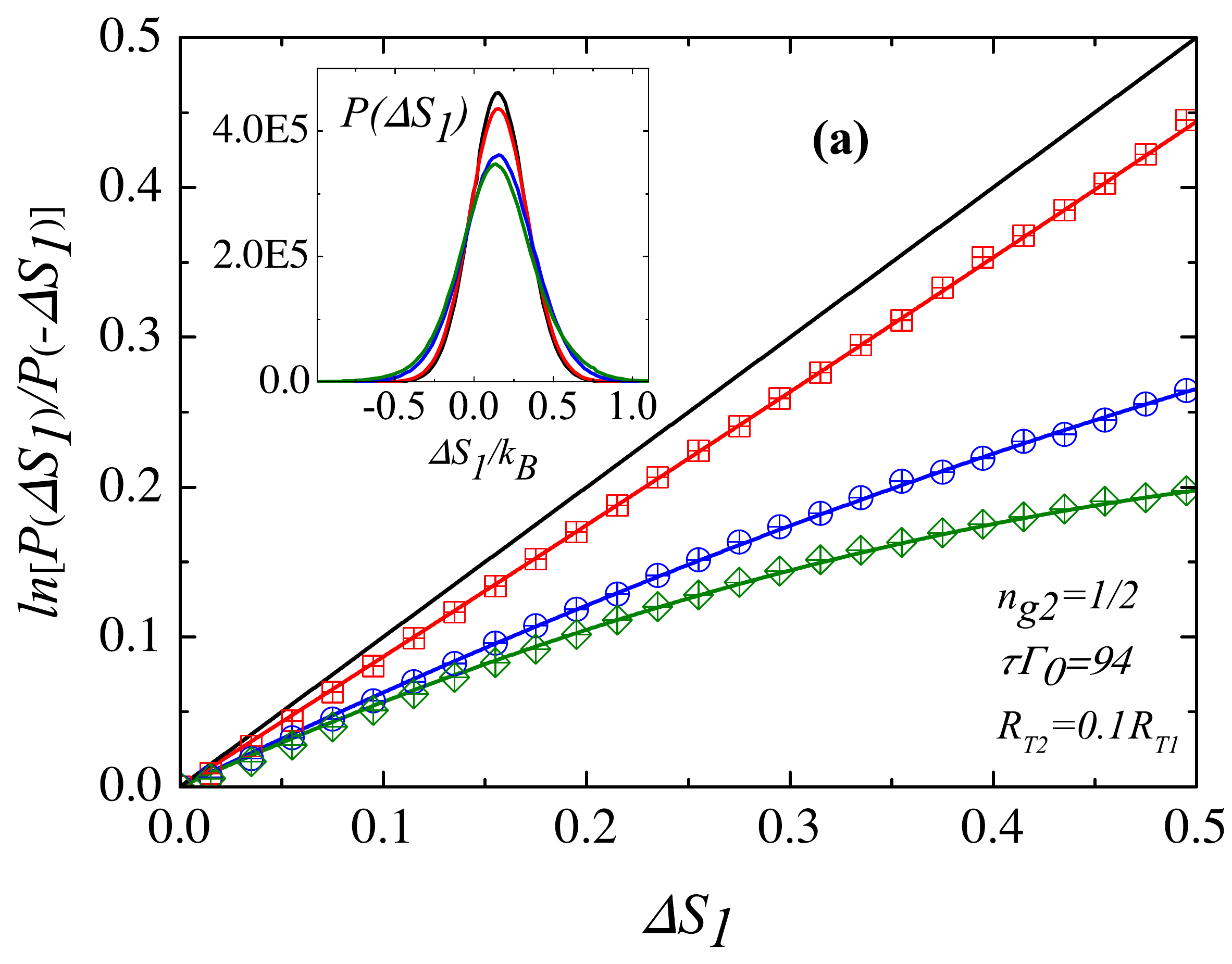}
\end{subfigure}
\begin{subfigure}{0.4\textwidth}
\centering
\includegraphics[width=\textwidth]{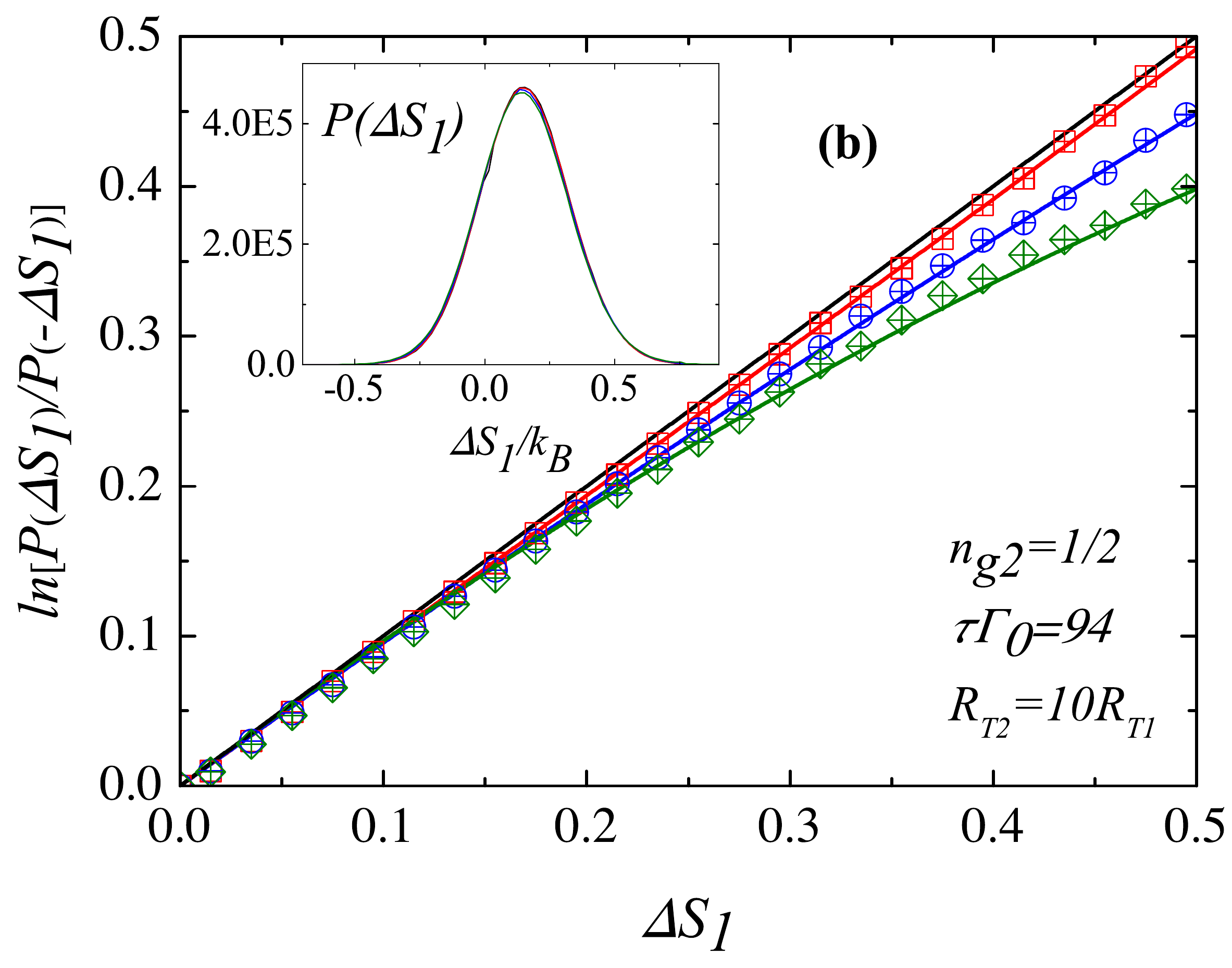}
\end{subfigure}
\caption{Crooks-type fluctuation relation for the single-box entropy production $\Delta S_1/k_{B}$ in the case of a slow drive $\tau\Gamma_0 =94$, for $R_{T_{2}}/R_{T_{1}}=0.1$ (a) and $R_{T_{2}}/R_{T_{1}}=10$ (b). The values of $J/E_{C}$ are the same as in Fig.~\ref{plot2} as well as the respective colors. Also, we show both the raw numerical data with symbols and their quadratic fitting with continuous lines. Inset: probability distribution of the single-box entropy production $P(\Delta S_{1})$.}
\label{plot3}
\end{figure}

We now let the two boxes to be unequal. More specifically we change the resistance $R_{T_{2}}$ of the second box relatively to the resistance $R_{T_{1}}$ of the first box. Since the transition rates are inversely proportional to the box resistance, by changing the ratio $R_{T_{2}}/R_{T_{1}}$ we are changing the typical dissipation time-scale of one box with respect to the other.\\ 
We first consider the situation where the resistance of the second SEB is smaller, leading to a larger transition rate. 
We might expect a behavior that deviates more strongly from Eq.~\eqref{totalTC} than that displayed in Fig.~\ref{plot2}a. By comparig the insets  of Fig.~\ref{plot2}a and Fig.~\ref{plot3}a we notice that in the latter all the distributions $P(\Delta S_{1})$ are slightly broader, especially on the negative side of the $\Delta S_{1}$-range. This implies that the probability of observing certain events increases with respect to the case of equal resistances. Since the only difference between the two models is a lower $R_{T_{2}}$ we are led to believe that these contributions stem from transitions in the second box that are more likely to happen even in the case of slow drive. However, such a difference is not immediate when we look at $\log\left[P(\Delta S_{1})/P(-\Delta S_{1})\right]$, Fig.~\ref{plot3}a. Here, the behavior is rather similar to the one shown in Fig.~\ref{plot2}a and no appreciable difference can really be noticed.\\
In the second case we instead assume the resistance of the second box to be larger. This translates to slowing down the second SEB with respect to the first one and it means that even for slow drive it should be harder to induce transitions in box 2 through the box-box interaction. Therefore, we might expect less participation of the second box to the total dynamics as well as to the measurement statistics. Figure~\ref{plot3}b confirms this prediction as, even with $\tau\Gamma_{0}=94$, the logarithm of the forward to backward distribution ratio $\log\left[P(\Delta S_1)/P(-\Delta S_1)\right]$ when the box-box interaction is on, deviates very little from the non-interacting case where we recover Eq.~\eqref{totalTC}. Thus, increasing the resistance of the second SEB is qualitatively similar to implementing a faster protocol. This can be further seen by direct comparison between the insets of Fig.~\ref{plot2}b and Fig.~\ref{plot3}b. Apart from the presence of no-transition peaks, the two sets of distributions look quite alike. Changing the interaction strength does not modify much the shape of the distributions and any deviations can be detected only when taking the logarithm of the forward to backward distribution ratio. Moreover, also in this case a non-linear trend appears only for higher values of $J$.\\
\section{Double drive}\label{double}
\begin{figure}[t!]
\centering
\includegraphics[scale=0.3]{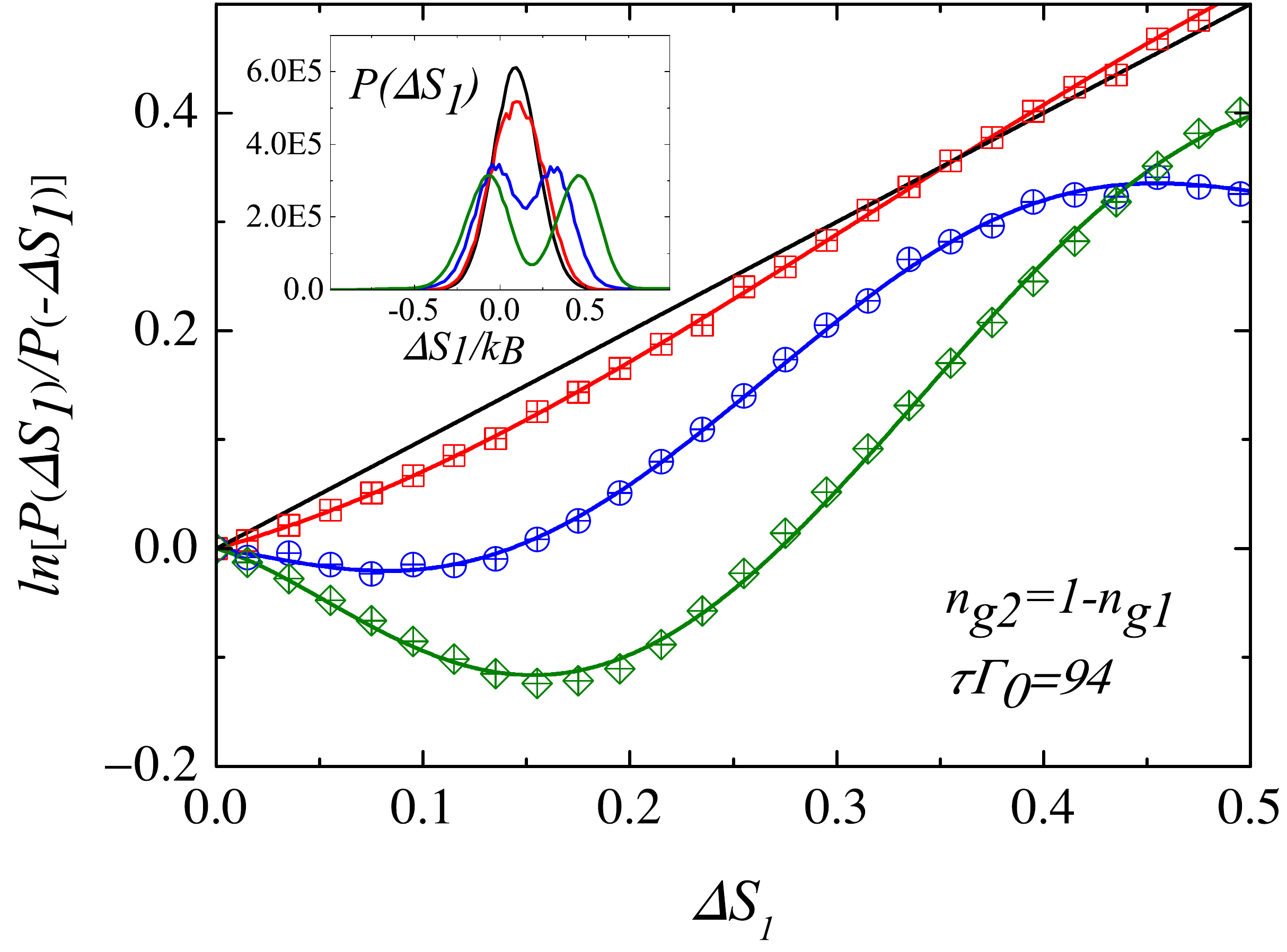}
\caption{Crooks fluctuation relation for the single-box entropy production $\Delta S_1/k_{B}$ in the case of a slow drive $\tau\Gamma_0 =94$, for the double-drive protocol. The values of $J/E_{C}$ are the same as in Fig.~\ref{plot2} as well as the respective colors. The continuous lines represent a $5^{\textrm{th}}$ order polynomial fitting of the raw numerical data (symbol). Inset: probability distribution of the single-box entropy production $P(\Delta S_{1})$.}
\label{plot4}
\end{figure}

\begin{figure}[h!]
\centering
\includegraphics[scale=0.3]{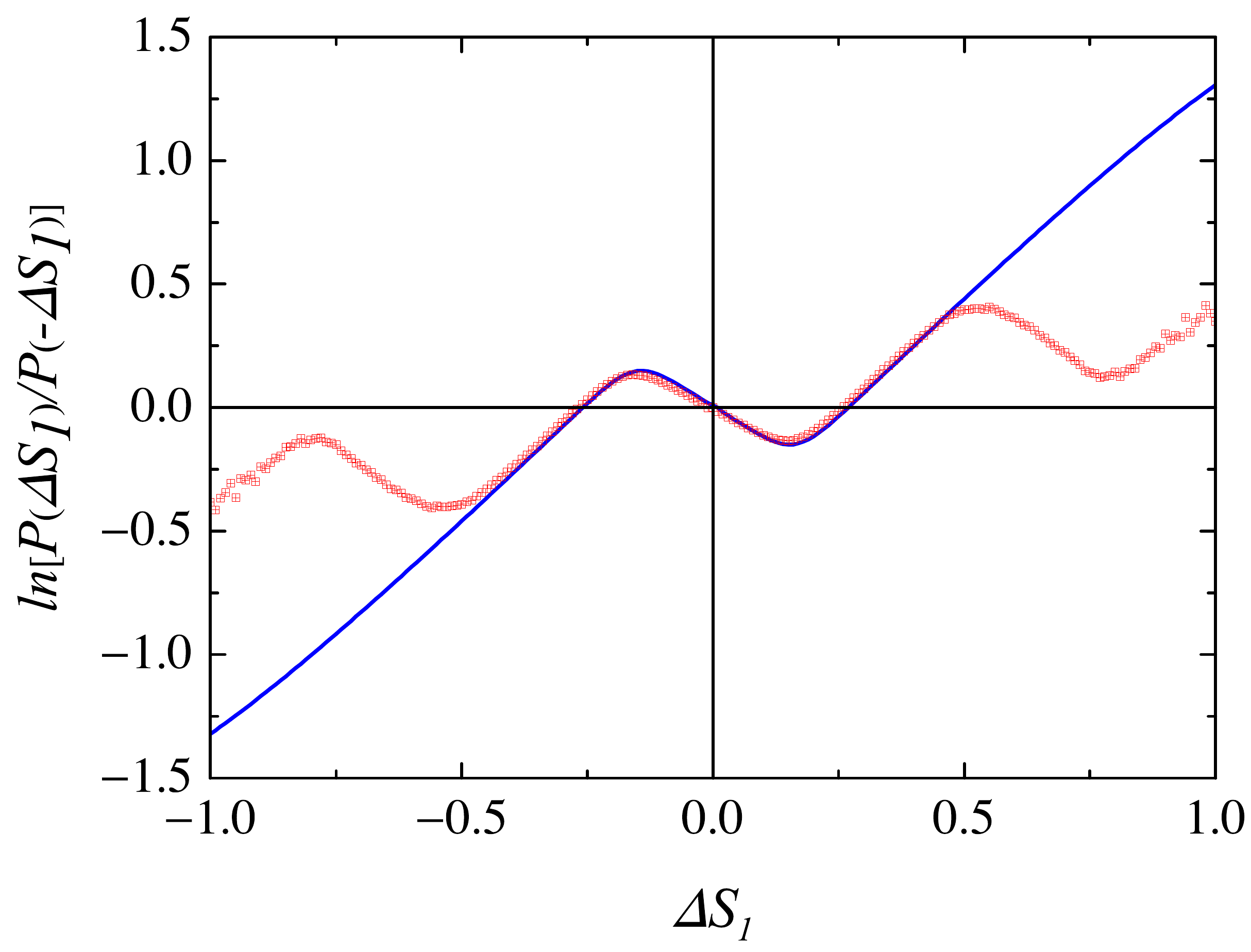}
\caption{Direct comparison between $\log\left[P(\Delta S_{1})/P(-\Delta S_{1})\right]$ as computed by Monte-Carlo simulations (red points) and from the single-jump analytical mode (blue line) for $\tau\Gamma_0 =94$ and $J/E_{C}=0.5$.}
\label{plot5}
\end{figure}

In this section we repeat the above study but with a different gate operation. We again assume the two SEBs to be at thermal equilibrium with the surrounding bath. While the first box is still subject to the same drive as in Eq.~\eqref{drive1} the second box gate voltage will be changed according to the following protocol
\begin{equation}
n_{g_{2}}(t)=1-n_{g_{1}}(t)=1-\frac{t}{\tau},
\label{drive2}
\end{equation}
see Fig.~\eqref{fig1}c. Thus, the two protocols are antisymmetric with respect to each other and if $n_{g_{1}}(t)$ drives the first box as $n_{1}:0\rightarrow1$, $n_{g_{2}}(t)$ will drive the second as $n_{2}:1\rightarrow0$. Since we consider identical boxes, the initial and final Hamiltonian are formally equal, leading to a vanishing free energy difference $\Delta F=0$. In this scenario extra work on the second box is performed and not accounted for. Therefore, we can expect modifications of the standard fluctuation relations to be more pronounced. 
 Although both boxes are being simultaneously driven, leading to what may seem  more complex dynamics, this particular choice of the total driving protocol results in some simplifications that allow us to perform analytical predictions. The ground state of the initial Hamiltonian is $(0,1)$ and therefore it is most probable in the initial Boltzmann distribution. Since the target process we want to realize is $(0,1)\rightarrow(1,0)$ and only single-electron processes are possible, we assume that most of the stochastic trajectories occurring are the following two-step trajectories 
\begin{equation}
\begin{aligned}
&\gamma_{1}:\; (0,1)\rightarrow(1,1)\rightarrow(1,0),\\
&\gamma_{2}:\; (0,1)\rightarrow(0,0)\rightarrow(1,0).
\end{aligned}
\label{2drivetraj}
\end{equation} 
We call this the single-jump approximation. By analyzing the corresponding transition rates, it is easy to see that such trajectories are equally probable. 
The total probability distribution of the entropy generated in the first SEB can be split in the two contributions arising from these trajectories
\begin{equation}
P(\Delta S_{1})=P_{\gamma_{1}}(\Delta S_{1})+P_{\gamma_{2}}(\Delta S_{1})
\label{ps1distranalyt}
\end{equation}
where $P_{\gamma_{1(2)}}(\Delta S_{1})$ is the probability distribution associated to the $\gamma_{1(2)}$ trajectory, whose analytical expressions are given in the Appendix. This model will be used in the following for a direct comparison with the numerical results, which we now report and discuss.
In Fig.~\ref{plot4} we show the the r.h.s. of Eq.~\eqref{TC1} for a slow driving protocol with $\tau\Gamma_{0}=94$ for several values of the box-box coupling parameter. Similarly to the previous cases, we display the values taken directly from simulations as well as their polynomial fittings, this time up to a fifth-order. Again the statistical confidence level is 0.95. A new feature we notice is the negativity of $\log\left[P(\Delta S_{1})/P(-\Delta S_{1})\right]$ for large coupling strengths and small entropy productions. 
What this negativity is telling us is that it is more probable for box 1 to lower its entropy rather than increase it during the execution of the transformation. This effect increases for increasing values of the magnitude of the interaction strength as indicated by an increase in the range of values for which $\Delta S_{1}$ is negative, up to a certain threshold after which positivity is restored. Although at first this result might appear bizarre and in contrast with the second law of thermodynamics, this is not the case. In fact, if we compute the average entropy production in box 1 we always find a positive quantity. Given the symmetry of the model, the same goes if we restrict our attention to the second SEB. As a matter of fact, the negativity arises as a consequence of the double-peak distribution shown in the inset. As mentioned above the two trajectories $\gamma_{1}$ and $\gamma_{2}$ occur with equal probability as clearly shown by the equal heights of the  peaks in $P(\Delta S_{1})$.
However, as soon as we turn the box-box interaction on, they become energetically unequal. This very asymmetry causes the negativity of $\log\left[P(\Delta S_{1})/P(-\Delta S_{1})\right]$ since the sum of the energies corresponding to single-box transitions equals the heat dissipated by the box itself which, in turn, is proportional to the entropy generated. Furthermore, since by increasing $J/E_{C}$ we increase the entropy gap between $\gamma_{1}$ and $\gamma_{2}$, a stronger intra-box coupling will display a progressively stronger negativity feature.\\
These results are obtained by statistically averaging over 10 million repetitions. From these data one can also reconstruct the joint occupation probabilities $p_{n_{1}n_{2}}(t)$ and check that they converge to the solutions of \eqref{RE}. Each repetition in the sample represents a single experiment and it is simulated via Monte-Carlo method with time-dependent transition rates where multiple back-and-forth transition trajectories are possible. One might then wonder the limits of applicability of the single-jump-trajectory approximation \eqref{2drivetraj}. In Fig. \ref{plot5} we compare $\log\left[P(\Delta S_{1})/P(-\Delta S_{1})\right]$ as obtained numerically (black) and analytically (red). For the sake of completeness the standard line on the r.h.s. of Eq.~ \eqref{totalTC} is displayed as well. The match between the two curves is  good up to $\Delta S_{1}\approx0.5$. 
From this point on the two curves no longer agree. While the single-jump model predicts a constant linear increase, with the Monte-Carlo method we observe a smooth decrease that appears almost sinusoidal. This feature is a consequence of multi-jump trajectories. Since the single-jump model only allows for just two electronic transitions during the execution of the whole protocol, it becomes less reliable at higher energies where electrons can tunnel many times back and forth in a single realization. This feature is imprinted in the local minima of the red curve. For instance, let us focus on the second minimum on the positive $\Delta S_{1}$ semi-axis, roughly at $\Delta S_{1}\approx0.9$. For this value of the entropy production the relevant trajectories have four electron jumps overall, three in one box and one in the other. The next local minima will be generated by further combinations of odd numers of single box electronic transitions summing up to an even number. This trend is observed for smaller values of $J/E_{C}$. Again, as soon as the second box is accounted for, no deviation from the standard behavior of Eqs.~\eqref{JarzQ} and \eqref{totalTC} is observed.
\section{Conclusions}\label{conc}
Partial observation of a system undergoing a non-equilibrium transformation can result in deviations from standard fluctuations relations.
In this manuscript, we have focused our attention on a common and easily implementable physical setup: a system of coupled single-electron boxes.  We have simulated two cases of experiment where one of the two SEBs is driven while the other one is either at rest (non-driven) or also subjected to work. The results indicate that $\log\left[P(\Delta S_{1})/P(-\Delta S_{1})\right]$ exhibits non-linear behavior. This is observed for a wide range of the two relevant parameters, duration of the protocol and strength of the box-box interaction. Furthermore, under some conditions, negativity can be observed in $\log\left[P(\Delta S_{1})/P(-\Delta S_{1})\right]$ as a consequence of a double-hump probability distribution of $\Delta S_{1}$. The intra-SEB interaction as well as both the driving protocols can be implemented with current technology, allowing for an experimental verification of the these results.
\section{Acknowledgments}
The authors would like to thank Dr. Ivan Khaymovich for useful discussions.
\section{Appendix I: single-jump approximation}
In this appendix we develop some tools to provide an analytical expression of the single-box entropy production probability  distribution under the assumption of single-jump trajectories. 
The relevant tunneling processes are
\begin{equation}\begin{aligned}
&1_{+}=(0,1)\rightarrow(1,1)\\
&2_{+}=(0,1)\rightarrow(0,0)\\
&3_{+}=(1,1)\rightarrow(1,0)\\
&4_{+}=(0,0)\rightarrow(1,0)
\end{aligned}
\label{subtransapp}
\end{equation}
and we label by $i_{-}$ the reverse processes. The energy releases $\Delta E_{i\pm}$ in each tunneling process, normalized by $E_C$ ($\Delta \epsilon_{i\pm}= \Delta E_{i\pm}/E_C$), read
\begin{eqnarray} \label{e2}
&\Delta \epsilon_{1+}= -\Delta \epsilon_{1-}=2n_{g1}-1-J(1-n_{g2})\nonumber \\&\Delta \epsilon_{2+}=-\Delta \epsilon_{2-}=1-2n_{g2}-Jn_{g1}\nonumber \\& \Delta \epsilon_{3+}= -\Delta \epsilon_{3-}=1-2n_{g2}+J(1-n_{g1})\nonumber \\& \Delta \epsilon_{4+}= -\Delta \epsilon_{4-}=2n_{g1}-1+Jn_{g2}.
\end{eqnarray}
For the chosen gate protocol $\Delta \epsilon_{1+}=\Delta \epsilon_{2+}$, $\Delta \epsilon_{3+}=\Delta \epsilon_{4+}$, and $\Delta \epsilon_{3+}=\Delta\epsilon_{1+}+J$.
We consider the single jump (in each box) trajectories of either $(0,1)\rightarrow (1,1)\rightarrow (1,0)$ or $(0,1)\rightarrow (0,0)\rightarrow (1,0)$ with equal probabilities, and with the total probability denoted $P_S$ which can be written in standard manner
\begin{widetext}
\begin{equation} \label{e4}
P_S = \int_0^\tau d\tau_2\int_0^{\tau_2} d\tau_1 e^{-\int_0^{\tau_1}[\Gamma_{1+}(\tau')+\Gamma_{2+}(\tau')]d\tau'}\big[\Gamma_{1+}(\tau_1)e^{-\int_{\tau_1}^{\tau_2}[\Gamma_{3+}(\tau')+\Gamma_{1-}(\tau')]d\tau'}\Gamma_{3+}(\tau_2)+\Gamma_{2+}(\tau_1)e^{-\int_{\tau_1}^{\tau_2}[\Gamma_{4+}(\tau')+\Gamma_{2-}(\tau')]d\tau'}\Gamma_{4+}(\tau_2)\big]
e^{-\int_{\tau_2}^\tau [\Gamma_{3-}(\tau')+\Gamma_{4-}(\tau')]d\tau'}.
\end{equation} 
\end{widetext}
The corresponding distribution of $q\equiv Q_1/E_C$ can be written as
\begin{equation} \label{e5}
P(q) = P_1(q)+P_2(q),
\end{equation}
where (in a form suitable for numerical implementation) we may write
\begin{widetext}
\begin{equation} \label{e6}
P_1(q)=r^2e^{-2r\int_{-1}^q \gamma_+(q')dq'}\gamma_+(q)
\int_q^{1-J} dq_2 e^{-r\int_q^{q_2} [\gamma_+(q'+J)+\gamma_-(q')]dq'}\gamma_+(q_2+J)e^{-2r\int_{q_2}^{1-J} \gamma_-(q'+J)dq'},
\end{equation}
\end{widetext}
and
\begin{widetext}
\begin{equation}
\label{e7}
P_2(q)= r^2e^{-2r\int_q^1 \gamma_-(q')dq'}\gamma_+(q)
\int_{-1+J}^q dq_1 e^{-2r\int_{-1+J}^{q_1} \gamma_+(q'-J)dq'}\gamma_+(q_1-J)e^{-r\int_{q_1}^{q} [\gamma_+(q')+\gamma_-(q'-J)]dq'}.
\end{equation}
\end{widetext}
Here, $r=E_C\tau/[(2-J)e^2R_T]$, and $\gamma_\pm (q)=\pm q/(1-e^{\mp\beta E_C q})$.

\end{document}